\documentclass{aa}
\usepackage{aas_macros}
\usepackage{natbib}
\usepackage{graphicx}
\usepackage{txfonts}

\def\wig#1{\mathrel{\hbox{\hbox to 0pt{%
          \lower.6ex\hbox{$\sim$}\hss}\raise.4ex\hbox{$#1$}}}}

\def\mearth{\rm M_\oplus}

\begin{document}

\title{A correlation between the heavy element content of transiting
extrasolar planets and the metallicity of their parent stars}
\titlerunning{The metallicity of extrasolar planets}
\authorrunning{Guillot et al.}

\author{Tristan Guillot\inst{1}
  \and
  Nuno C. Santos\inst{2,3,4}
  \and
  Frederic Pont\inst{3}
  \and
  Nicolas Iro\inst{5}
  \and
  Claudio Melo\inst{6}
  \and 
  Ignasi Ribas\inst{7}
  }

\offprints{T. Guillot}

\institute{Observatoire de la C\^ote d'Azur, CNRS UMR 6202, 06304 Nice
 Cedex 4, France
	\and Centro de Astronomia e Astrof\'{\i}sica da Universidade de
  Lisboa, Observat\'orio Astron\'omico de Lisboa, Tapada da Ajuda,
  1349-018 Lisboa, Portugal
	\and Observatoire de Gen\`eve, 51 ch. des Maillettes, CHÐ1290
  Sauverny, Switzerland
  \and Centro de Geofisica de Evora, Rua Romeo Ramalho 59,
 7002-554 Evora, Portugal
	\and Department of Physics, University of Florida, PO Box 118440,
  Gainesville, FL 32611-8440, USA
	\and European Southern Observatory, Casilla 19001, Santiago 19, Chile 
	\and Institut d'Estudis Espacials de Catalunya - CSIC, Campus
 UAB, Facultat de Ci\`encies, Torre C5-parell-2a, 08193 Bellaterra,
 Spain}

\date{Accepted for publication in A\&A, April 30, 2006.}

\abstract{
  Nine extrasolar planets with masses between 110 and 430$\mearth$
  are known to transit their star. The knowledge of their masses and
  radii allows an estimate of their composition, but uncertainties on
  equations of state, opacities and possible missing energy sources
  imply that only inaccurate constraints can be derived when considering
  each planet separately.}
{We seek to better understand the composition of transiting
  extrasolar planets by considering them as an ensemble, and by
  comparing the obtained planetary properties to that of the
  parent stars.}
{We use evolution models and constraints on the stellar ages to
  derive the mass of heavy elements present in the
  planets. Possible additional energy sources like tidal dissipation
  due to an inclined orbit or to downward kinetic energy transport are
  considered.}
{We show that the nine transiting planets discovered so far belong to
  a quite homogeneous ensemble that is characterized by a mass of
  heavy elements that is a relatively steep function of the stellar
  metallicity, from less than 20 earth masses of heavy elements around
  solar composition stars, to up to $\sim 100\,\mearth$ for three
  times the solar metallicity (the precise values being
  model-dependant). The correlation is still to be ascertained
  however. Statistical tests imply a worst-case 1/3 probability of 
  a false positive.}
{Together with the observed lack of giant planets in close orbits
  around metal-poor stars, these results appear to imply that heavy
  elements play a key role in the formation of close-in giant planets.
  The large masses of heavy elements inferred for planets orbiting
  metal rich stars was not anticipated by planet formation models and
  shows the need for alternative theories including migration and
  subsequent collection of planetesimals.} 
   

\maketitle
%
%________________________________________________________________

\section{Introduction}

Transiting extrasolar planets are extremely interesting objects to
study because the coupling of radial velocity and photometric
measurements allows a determination of both their masses and radii, and
thus, in principle, a constraint on their composition. For that
purpose, we need to know as accurately as possible the
properties of the star-planet systems, and to apply evolution models
using up-to-date input physics. 

Nine transiting extrasolar planets are known to date. In Table~1, we
summarize their characteristics and that of their parent stars on the
basis of available literature, including recently derived
metallicities and stellar effective temperatures from high-resolution
and high signal-to-noise spectra obtained with UVES for the faint
stars \object{OGLE-TR-10}, \object{56}, \object{111}, \object{113} and
\object{TrES-1} \citep{Santosetal2006} and \object{OGLE-TR-132}
\citep{Pontetal2006}. The ages were derived from isochrone fitting
when possible. In some cases, lower limits were obtained from lithium
abundances, the Ca II activity-age relation, and an analysis of the
stellar rotational velocity \citep{Meloetal2006}. In the absence of
useful constraints, we used a conservative upper limit of 10\,Ga
because the stars' metallicities and proper motions imply that they
all belong to the galactic thin disk.

%%
%% TABLES
%%
\begin{table*}
  \caption{Characteristics of the transiting Pegasi planets discovered
  so far} 
  \label{tab:transits}
%  \small
  \vspace*{-0.3cm}
  \begin{tabular}{llrrrrrrr} \hline \hline
\# & Name & Age [Ga] & [Fe/H] & $T_{\rm eff}$ [K] & $T_{\rm eq,0}$ [K]
& $M_{\rm p} [\rm M_{\oplus}]$ &  $R_{\rm p}$ [Mm] & Refs. \\
1 & {\bf \object{HD209458}} & $4-7$ & $0.02(3)$ & $6117(30)$ & $1487(120)$ & $210(19)$ & $97(5)$ &
 [1,2] \\
2 & {\bf \object{OGLE-TR-56}} & $2-4$ & $0.25(8)$ & $6119(60)$ & $2079(140)$ & $394(73)$ & $88(11)$ & 
 [3,4,5] \\
3 & {\bf \object{OGLE-TR-113}} & $0.7-10$ & $0.15(10)$ & $4804(110)$ & $1340(80)$
& $429(70)$ & $77(5)$ & [5,6] \\
4 & {\bf \object{OGLE-TR-132}} & $0.5-2$ & $0.37(7)$ & $6210(59)$ &
$1980(100)$ & $350(38)$ & $80\left(^{+9}_{-4}\right)$ & [6,7,8] \\
5 & {\bf \object{OGLE-TR-111}} & $1.1-10$ & $0.19(7)$ & $5044(80)$ & $1033(160)$
& $168(35)$ & $71(4)$ & [3,5,9] \\
6 & {\bf \object{OGLE-TR-10}} & $1.1-5$ & $0.28(10)$ & $6075(90)$ & $1578(50)$ &
$200(41)$ & $102(7)$ & [3,4,5,10] \\
  &            &           &                &      &      &      &
$81(6)$ & [11] \\
7 & {\bf \object{TrES-1}} & $2-6$ & $0.06(5)$ & $5226(40)$ & $1157(140)$ &
$238(22)$ & $74\left(^{+6}_{-4}\right)$ & 
[5,12] \\
8 & {\bf \object{HD149026}} & $1.2-2.8$ & $0.36(5)$ & $6147(50)$ & $1740(150)$ & $114(10)$ & $52(4)$ &
[13,14] \\
9 & {\bf \object{HD189733}} & $0.5-10$ & $-0.03(4)$ & $5050(50)$ & $1199(30)$ & $365(13)$ & $90(2)$ & 
[15] \\
  &            &           &                &      &      &      &
$82.5(23)$ & [16] \\
    \hline \hline
  \end{tabular}

  The numbers in parenthesis represent the uncertainties on the
  corresponding last digits.\\
  References: [1] \citet{Winnetal2005}; [2] \citet{Santosetal2004};
  [3] \citet{Udalskietal2002}; [4] \citet{Bouchyetal2005a}; [5]
  \citet{Santosetal2006}; [6] \citet{Bouchyetal2004}; [7]
  \citet{Moutouetal2004}; [8] \citet{Pontetal2006}; [9]
  \citet{Pontetal2004}; [10] \citet{Konackietal2005}; [11]
  \citet{Holmanetal2006}; [12] \citet{Sozzettietal2004}; [13]
  \citet{Satoetal2005}; [14] \citet{Charbonneauetal2006}; [15]
  \citet{Bouchyetal2005b}; [16] \citet{Bakosetal2006}. 
  \normalsize
\end{table*}

The radii of close-in extrasolar giant planets (hereafter Pegasi
planets or Pegasids, after \object{51 Peg b}) of a given composition
will depend on essentially three quantities: their masses, ages, and
the amounts of flux they receive from their parent star. The latter
are expressed in Table~1 in the form of $T_{\rm eq,0}$, the effective
equilibrium temperature calculated for a zero planetary albedo. In
terms of these parameters, the sample of known transiting Pegasi
planets is already quite rich. And indeed, evolution calculations of
individual planets have yielded quite different results. The first one
discovered, \object{HD209458b}, was shown to be anomalously large
\citep{Bodenheimeretal2001,GuillotShowman2002}. The subsequent six
planets appeared relatively "normal", i.e. fitting the standard
evolution models within the error bars, with the exception of
\object{OGLE-TR-10b}, another too-large planet
\citep{Baraffeetal2005,Laughlinetal2005}. Then, \object{HD149026b} was
shown to be significantly smaller than expected, requiring the
presence of a large amount $\sim 70\mearth$ of heavy elements in its
interior \citep{Satoetal2005}. The last addition to that list is that
of \object{HD189733b} \citep{Bouchyetal2005b}, which, like
\object{HD209458b} orbits a star of near-solar metallicity, and is
found to be relatively large.

While a radius that is smaller than expected by theoretical evolution
calculations for a solar-composition giant planet can be easily
accounted for by the presence of a central dense core, or generally
more heavy elements in the interior, large radii point towards missing
physics in the models: an additional energy source, or inaccurate
physical inputs (equations of state, opacities, atmospheric
models). It has been proposed that the anomalous radius of HD209458b
could be explained by tidal heating linked to a small forced
eccentricity $e\sim 0.03$ \citep{Bodenheimeretal2001}, but detailed
observation indicate that the eccentricity is small, $e=0.014\pm
0.009$ \citep{Laughlinetal2005b}, and observations of the secondary
eclipse imply that this would further require a chance configuration
of the orbit \citep{Demingetal2005}. Another proposed explanation also
involving tidal dissipation of orbital energy is that the planet may
be trapped in a Cassini state with a large orbital inclination
\citep{WinnHolman2005}. Finally, a third possibility that would apply
to {\it all} Pegasi planets is to invoke a downward transport of
kinetic energy and its dissipation by tides
\citep{ShowmanGuillot2002}. This last possibility would require the
various transiting planets to have different core masses to reproduce
the observed radii \citep{Guillot2005}.

The purpose of this article is to test whether these scenarios are
possible, and to attempt to constrain the masses of heavy elements
present in these planets. We first compare measured radii to the ones
obtained by standard evolution models of extrasolar planets. We derive
in Section~3 masses of heavy elements and compare them to the
metallicities of the parent stars. Consequences for our understanding
of planet formation are then discussed.

\section{Standard models and the radius anomaly}

\begin{figure}
\centerline{\resizebox{10cm}{!}{\includegraphics{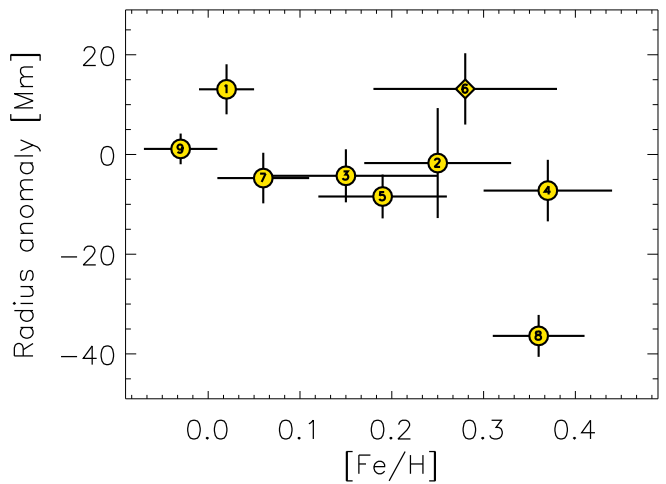}}}
\caption{Radius anomaly (in units of Mm, or 1000 km) of the known
  transiting Pegasi planets as a function of the metallicity of their
  parent star (see Table~1 for a list of the planets). The radius
  anomaly is defined as the difference between the measured radius and
  the one calculated using a simple evolution model for a solar
  composition gaseous planet. }
\end{figure}

We first calculate evolution models for all nine planets based on a
simple model assuming a near-solar composition \citep{Guillot2005}. We
use a pure hydrogen-helium equation of state
\citep{SaumonChabrierVanHorn1995} with a helium mass mixing ratio
$Y=0.30$ and standard Rosseland opacities \citep{Allardetal2001}. The
atmospheric boundary condition is calculated for each planet for an
incoming stellar flux averaged over the entire planetary surface
\citep{Iroetal2005}. We stress that significant uncertainties on the
atmospheric temperature profile arise from incomplete opacity sources,
presence or absence of clouds, horizontal inhomogeneities, and
advective transport of heat. This implies that atmospheric
temperatures in the literature can vary by up to 500\,K
\citep{Barmanetal2001,Sudarskyetal2003,
Fortneyetal2005,Iroetal2005,Seageretal2005}. For our purposes however,
this will not matter as long as the temperature of the deep atmosphere
remains roughly proportional to the equilibrium temperature.  For
example, in our calculations of atmospheric profiles, the ratio of the
temperature at the 10 bar level to the equilibrium temperature is
found to be between 1.18 and 1.28 for all the planets in Table~1. 
%The
%robustness of the results presented hereafter was successfully tested
%with values of this parameter between 1 and 1.5.

We define the {\it radius anomaly} as the difference between the
measured radius and the one obtained by our evolution model for the
ages inferred in Table~1, and accounting for the fact that the
measured radii correspond to $\sim $mbar levels
\citep{Burrowsetal2003,Fortneyetal2003,Iroetal2005}. Figure~1 shows
the radius anomaly thus obtained against the metallicity of the parent
star. The error bars on the radius anomaly are equivalent $1\sigma$
error bars calculated on the basis of gaussian distribution of radii
measurements and on a uniform distribution of the age within the
limits in Table~1. (Note that for simplicity, in the case of
\object{OGLE-TR-132b} and \object{Tres-1b}, we arbitrarily recentered
the radius measurement to have symetrical error bars). Figure~1
indicates that the radius anomaly and the metallicity of the parent
star may be anticorrelated.
% At this point, the case for this
%anticorrelation is weak, as it is mainly driven by the two extrema for
%HD209458b and HD149026b. 
We found no such correlation between the radius anomaly and the age,
mass, effective temperature of the star, or with the mass of the
planet, and equilibrium temperature. We also found no correlation with
the total XUV flux received during the planets' lifetime, implying
that planetary evaporation probably remained limited in magnitude.

In figure~1 and hereafter, two planets deserve a special discussion:
\object{OGLE-TR-10b} and \object{HD189733b} are characterized by two
significantly different radius determinations (see table~1). In the
case of \object{OGLE-TR-10b}, we adopted the value by
\citet{Bouchyetal2005a}, because the value by \citet{Holmanetal2006}
remains unconfirmed thus far. A special symbol for
\object{OGLE-TR-10b} is used to remind us of this uncertainty. For
\object{HD189733b}, the most recent value \citep{Bakosetal2006} was
adopted because the photometric analysis appears to be improved over
the old value. In both cases, this corresponds to a worst-case
situation for the (anti)correlation that we are seeking.

The fact that the radius anomaly decreases with increasing [Fe/H]
indicates that the source of the correlation is not in the atmosphere
because larger amounts of heavy elements would lead to larger
atmospheric temperature \citep{Fortneyetal2005} and therefore larger
radii. On the other hand, the magnitude of the anomaly is much too
large to be explained by assuming that the planets have the same
composition as their stars. On this basis we now attempt to infer the
masses of heavy elements present in the planets.

\section{Masses of heavy elements}

At this point, the correlation is just indicative however because the
radius anomaly also depends on the planetary mass. In order to
ascertain its reality and to constrain the planet composition/star
metallicity relationship, we further calculated evolution models for
planets with a central rock core and a solar-composition envelope
\citep[see][]{Guillot2005}. The results are shown in Figure~2, in
which we plot as a function of the metal content of the central star,
the amount of heavy elements that is required to reproduce the
observed planetary radii. In most cases, this amount is positive, but
for \object{HD209458b} and \object{OGLE-TR-10b}, the large measured
radii yield by extrapolation unphysical negative values of the mass of
heavy elements.

\begin{figure}
\centerline{\resizebox{10cm}{!}{\includegraphics{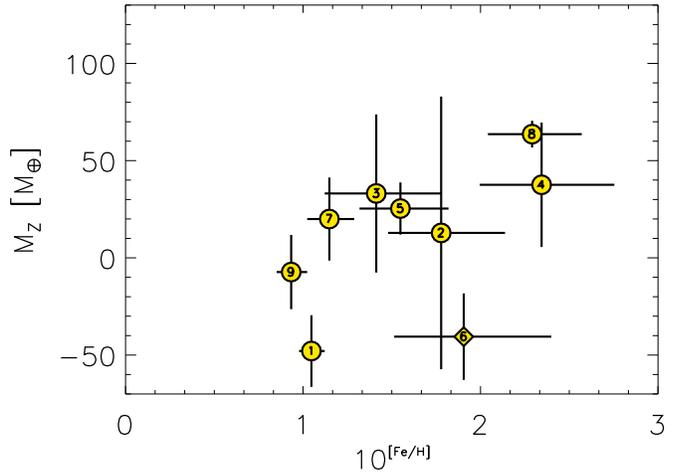}}}
\caption{Mass of heavy elements in the planets as a function of the
  metal content of the parent star relative to the Sun. The mass of
  heavy elements required to fit the measured radii is calculated on
  the basis of standard evolution models. Negative core masses are
  required in some cases, implying that some significant physical
  input is missing (see text). Horizontal error bars correspond to the
  $1\sigma$ errors on the [Fe/H] determination. Vertical error bars
  are a consequence of the uncertainties on the measured planetary
  radii and ages.}
\end{figure}

One possible interpretation is that energy dissipation occurs only in
\object{HD209458b} and \object{OGLE-TR-10b} (e.g. via tidal
dissipation due to a non-zero inclination of the orbital plane), which
could explain the negative core masses derived for these planets.  If
we retain this explanation, the remaining 7 planets with positive core
masses appear to possess amounts of heavy elements that are correlated
with the stellar metallicity. The corresponding linear correlation
coefficient for these 7 points is then relatively high, $r=0.78$
(Pearson) or $r_{\rm S}=0.71$ (Spearman), with a corresponding
significance level (no-correlation probability) close to 7\%. A tough
test is however to retry all the points within their error bars and
calculate an averaged significance level, which we find to be $\sim
33\%$.

An alternative explanation that we favor is that some physics is
missing in the treatment of the evolution calculation and affects {\em
  all} the planets. One possibility could be that the hydrogen
equation of state severely underestimates the density for a given
pressure and temperature. It could also be due to a strongly
underestimated deep opacity that would then artificially shorten the
planetary cooling and contraction. But given that it would require
relatively severe modifications to these physical inputs, we will only
focus here on a third possibility: that downward kinetic energy
transport and its dissipation due to tidal effects in the planetary
interior effectively leads to an additional energy source that slows
the contraction of the planets
\citep{GuillotShowman2002,ShowmanGuillot2002}. Quantitatively, we
found that adding an additional energy source at the planet center equal
to $0.5\%$ of the incoming stellar luminosity was sufficient to
solve the problem of \object{HD209458b} and \object{OGLE-TR-10b}.

Figure~3 shows the masses of heavy elements that result from these
evolution models including this additional energy source. A pleasing
feature of these models is that all planets can be explained within
the same physical framework, and that the planet-star metallicity
correlation appears to be reinforced: including all nine planets, the
correlation coefficients are $r=0.58$, $r_{\rm S}=0.50$, for a
significance level of 17\%, which is confirmed by permutation and
bootstrap tests. Using Monte-Carlo retrials of the nine points within
their error bars, the mean significance level increases to
$27\%$. Removing the questionned \object{OGLE-TR-10b} point from the
sample, the rank correlation coefficient increases to $r_{\rm
S}=0.83$, for a false positive probability of $\sim 1\%$ and a mean
significance level of the Monte-Carlo retrials of 12\%. Other
calculations (not presented) with a higher fraction of dissipated
energy (1\% of the incoming flux), or with an ad-hoc no-helium
equation of state also lead qualitatively to the same results.

\begin{figure}
\centerline{\resizebox{10cm}{!}{\includegraphics{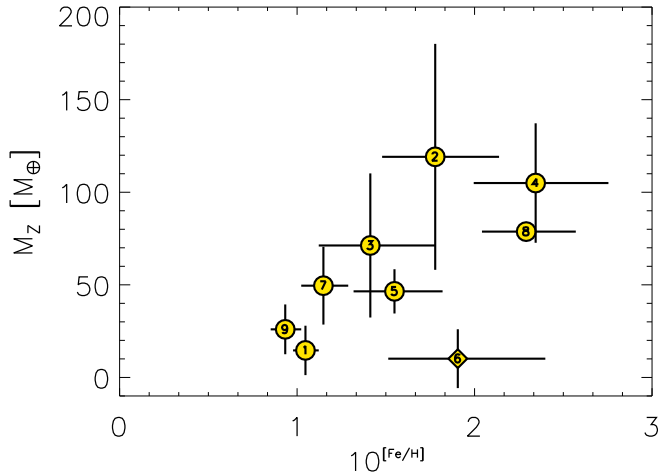}}}
\caption{Same as fig.~2, but the mass of heavy elements required to
  fit the measured radii are calculated on the basis of evolution
  models including an additional heat source slowing the cooling of
  the planet. This heat source is assumed equal to $0.5\%$ of the
  incoming stellar heat flux \citep{ShowmanGuillot2002}.}
\end{figure}

\section{Conclusions}

We have shown that the nine transiting extrasolar planets discovered
so far, including the anomalously large \object{HD209458b} and the
``big core'' planet \object{HD149026b} form a relatively homogeneous
ensemble, whose compositions can be explained by the same evolution
model for {\em all} planets. Alternatively, we cannot rule out that
the structure of one or several planets was affected by particular
physical processes (e.g. a non-zero inclination giving rise to tides,
a different orbital evolution...etc.).

We found that the masses of heavy elements in the planets appear to be
proportional to the metallicities of their parent star. This
correlation remains to be tested, being still consistent with a
no-correlation hypothesis at the 1/3 level in the least favorable
case. If true, it is relatively steep ($M_Z\sim 60\,{\rm
M_\oplus}(10^{[Fe/H]}-0.8)$) and appears to exclude the presence of
Pegasi planets around stars with a metallicity smaller than that of
the Sun by more than $\sim 0.1$dex, in agreement with the lack
of Pegasids ($a<0.1$ AU) around metal-poor stars observed by
radial velocimetry.  

The steepness of the relation, and the large values of $M_Z$ obtained
in some cases are surprising and have not been predicted by formation
models, either in the framework of the core-instability model
\citep{IdaLin2004,Alibertetal2005} or by models involving direct
gravitational instability of the gas \citep{Boss2003}. The core
instability model seems however to provide the key ingredients to
explain this relation: a faster growth of solid planetary embryos with
increased metallicity, the subsequent collection of these during or
after the capture of a gaseous envelope, and a more likely migration
for planets that are formed quickly. On the other hand, in the
framework of gravitational instability models, one needs to assume
that the gaseous protoplanets are able to collect dust quickly and
that the efficiency of this process is directly linked to the
metallicity of the disk. However, we would then expect to find planets
around metal-poor stars which is not observed both by transit surveys
(although figs.~2 and 3 imply they would be easier to detect) and by
radial velocimetry. It seems therefore unlikely that gravitational
instability is responsible for the formation of Pegasi planets.

Another consequence of this work is that the most typical Pegasids,
$150\rm\,M_{\oplus}$ planets orbiting stars more metal-rich than the
Sun, will be smaller than expected. Since the transit depth scales
with the square of the radius, this has tough implications for
ground-based transit searches.  For instance, a $60$\,Mm ($0.86\,\rm
R_J$) planet would produce a $0.7\%$ transit in front of a
$1\rm\,R_\odot$ star, instead of $1.2\%$ for a $80$\,Mm ($1.15\,\rm
R_J$) planet as typically assumed, enough to sink it below the
detection threshold of all present ground-based surveys.  And indeed,
the planets found around the three most metal-rich stars by
photometric transit searches, \object{OGLE-TR-132}, \object{10}, and
\object{56} (see Table~1) are also the ones that endure the most
irradiation from their parent star: they thus have been prevented from
contracting below the threshold level. This may contribute to explain
the relative lack of success of photometric transit searches compared
to radial velocity surveys.

Clearly, although tantalizing, the results presented in this study are
limited by the relatively small number of known transiting planets. In
the next few years, the discovery of transiting extrasolar planets, in
particular with the space missions COROT (launch 2006) and Kepler
(launch 2008), their careful characterization coupled to studies of
their parent stars should greatly improve our understanding of planet
formation.

\bigskip\noindent
{\it Note added in proof}: 
The transiting planet recently discovered by \cite{McCullough2006},
\object{XO-1b}, is very similar in mass, radius and stellar
metallicity to HD209458b and further strengthens the correlation
proposed in this paper.

\bibliography{correlation.bib}

\end{document}